\documentclass{webofc}
\usepackage[varg]{txfonts}   % Web of Conferences font
\begin{document}
\title{Strangeness in Neutron Stars}

\author{\firstname{Laura} \lastname{Tolos}\inst{1,2,3}\fnsep\thanks{\email{tolos@ice.csic.es}} }

\institute{ 
Institute of Space Sciences, Campus UAB,  Carrer de Can Magrans, 08193 Barcelona, Spain
\and
Institut d'Estudis Espacials de Catalunya, 08034 Barcelona, Spain
\and
Frankfurt Institute for Advanced Studies, Ruth-Moufang-Str. 1, 60438 Frankfurt am Main, Germany          }

\abstract{%
In this contribution I briefly review the dynamics of strange mesons and baryons with  dense nuclear matter, paying a special attention to their presence in the inner core of neutron stars and the consequences for the structure of these compact stars.}
\maketitle
\section{Introduction}
\label{intro}
Understanding the dynamics of hadrons with strangeness has received a lot attention over the past decades, specially in connection with the investigation of the possible strange phases in the interior of neutron stars (see for  Ref.~\cite{Tolos:2020aln,Burgio:2021vgk} for recent reviews).

On the one hand, the dynamics of strange baryons (hyperons) with nuclear matter has been object of high interest due to the probable existence of hyperons in neutron stars, leading to the so-called hyperon puzzle \cite{Chatterjee:2015pua,Vidana:2018bdi}. On the other hand, the properties of strange mesons (antikaons)  in dense nuclear matter have prompted the interest of the scientific community, in particular after the predictions of kaon condensation in neutron stars \cite{Kaplan:1986yq}.

In this contribution I briefly review the field of strange hadrons (hyperons and antikaons) in neutron stars, discussing the consequences for the structure of these compact objects. 

\section{ Neutron Stars: masses, radii and tidal deformability}

Neutron stars are one of the most compact known stellar objects and, thus, turn out to be a quite unique laboratory for testing matter under extreme conditions of density \cite{Watts:2016uzu,Watts:2018iom}.  With typical masses of 1-2${\rm M_{\odot}}$ and radii in the range of 10-12 km, the average density within neutron stars is of the order of $\sim 10^{14}$ ${\rm g/cm^3}$. 

The mass and radius of these compact stars depend on the properties of matter in their interior and, thus, are strongly correlated to the equation of state of dense matter. Very accurate observations of 2${\rm M_{\odot}}$ neutron stars have been determined in the past decade \cite{Demorest:2010bx,Antoniadis:2013pzd,2020NatAs...4...72C} by means of measuring certain post-Keplerian parameters that account for deviations from the Keplerian orbit due to general relativistic effects in such compact objects. As for the radii, in the past they have been extracted from the analysis of X-ray spectra emitted by the neutron star atmosphere. These measurements of radii were rather difficult as the X-ray spectra strongly depend on the distance to the source, its magnetic field and the composition of its atmosphere. With space missions such as NICER (Neutron star Interior Composition ExploreR), high precision X-ray astronomy, based on pulse-profile modelling X-ray spectral-timing event data, is already offering precise measurements of masses and radii simultaneously \cite{Riley:2019yda,Miller:2019cac,Riley:2021pdl,Miller:2021qha}. 

Moreover, a new era in astrophysics has begun with the measurement of gravitational waves emitted from binary neutron star mergers, detected by the LIGO and VIRGO Collaborations \cite{LIGOScientific:2017vwq,LIGOScientific:2018hze}. Gravitational waves from the late inspirals of neutron stars are sensitive to the equation of state, through the so-called tidal deformability of the star, as the tidal effects are strongly dependent on the stellar compactness. Therefore, the tidal deformability allows discriminating among equations of state that predict similar masses but different radii.

\section{Hyperons inside Neutron Stars: The Hyperon Puzzle}
\label{hyperons}

The composition of the core of neutron stars is determined by imposing equilibrium against weak interaction processes, the so-called $\beta$-stability. Traditionally, the core of neutron stars has been modelled by a uniform fluid of neutron rich matter in  $\beta$-equilibrium. However, other degrees of freedom could be expected, such as hyperons. Hyperons might be energetically favoured due to the high value of density in the interior of neutron stars and the rapid increase of nucleon chemical potential with density.

The appearance of hyperons in the core of neutron stars would affect the structure of these compact stars. Indeed, the pressure becomes softer with respect to the case when only nucleons are present. This is due to the fact that the addition of one specie opens a set of new available low-energy states that can be filled, hence lowering the total energy (and pressure) of the system. 
When the total pressure of the star is lowered, the gravitational pull is reduced so as to keep the hydrostatic equilibrium. Therefore, the less pressure, the less mass the neutron star has, leading to masses below the $2M_{\odot}$ neutron star observations. This fact is often referred as {\it the hyperon puzzle}. 
 
Several scenarios  have been devised in order to solve this puzzle. These are: i) the use of stiffer hyperon-nucleon and hyperon-hyperon interactions; ii) the stiffening induced by hyperonic three-body forces; iii) the appearance of new hadronic degrees of freedom that push the hyperon onset to higher densities, such as $\Delta$ baryons or meson condensates; iv)  the presence of an early phase transition to quark matter below the hyperon onset; and v) the use of modified gravity models to accommodate hyperons in 2${\rm M_{\odot}}$ neutron stars. For reviews of these different scenarios see  Refs.~\cite{Vidana:2018bdi,Tolos:2020aln} and references therein.

\section{Antikaons in Neutron Stars: Kaon Condensation}
\label{kaons}
Another possible scenario inside the core of neutron stars is the appearance of antikaons. 
As mentioned in the previous section, the composition of matter in neutron stars is determined by demanding equilibrium against weak interaction processes. Thus, considering neutron star matter made of neutrons ($n$), protons ($p$) and electrons ($e$), the weak interaction transitions that take place are  $n \rightarrow p e^- \bar \nu_e$ and  $e^- p \rightarrow n \nu_e$, so that the chemical potentials of the different species are related by $\mu_n=\mu_p+\mu_e$, where the densities $\rho_p=\rho_e$, with the total baryonic density $\rho=\rho_p+\rho_n$. However, if the electron chemical potential  increases dramatically with density in the interior of neutron stars, it could be energetically more favourable to produce antikaons instead of electrons via  $n \leftrightarrow p + \bar{K}$ . In that case, the electron chemical potential  for a given density should be larger than the effective mass of antikaons at that density, that is, $\mu_e > m^*_{\bar K}$. If this was the case and given that antikaons are bosons,  {\it kaon condensation} would take place.

The possibility of kaon condensation has been put forward since the pioneer work of Ref.~\cite{Kaplan:1986yq}.  The question is whether the mass of antikaons could be largely modified by the interaction with the nuclear medium. Some phenomenological models tend to favour this scenario, specially those based in relativistic mean-field models (see, for example, the recent works  \cite{Gupta:2013sna,Thapa:2021kfo,Malik:2021nas,Muto:2021jms}). Nevertheless, microscopic unitarized schemes do not support this idea, since a large modification in the mass of antikaons is required for densities inside neutron stars and this is not obtained in these approaches. I refer the reader to Refs.~\cite{Lutz:1997wt,Ramos:1999ku,Tolos:2000fj,Tolos:2002ud,Tolos:2006ny,Lutz:2007bh,Tolos:2008di,Cabrera:2014lca} for the mass modification of antikaons in microscopic unitarized approaches based on meson-exchange models or chiral effective Lagrangians.

\section{Summary}

I have briefly reviewed the dynamics of strange hadrons (hyperons and antikaons) in the interior of neutron stars. On the one hand, I have discussed the probable presence of hyperons inside neutron stars, and  the so-called hyperon puzzle and its possible solutions. On the other hand, I have presented the phenomenon of kaon condensation in neutron stars and the controversy that surrounds it.  

\section*{Acknowledgments}

This contribution is based on an invited plenary talk at the 17th International Workshop on Meson Physics conference (Meson2023), sponsored by the European Physical Journal A. I also acknowledge support from CEX2020-001058-M (Unidad de Excelencia ``Mar\'{\i}a de Maeztu"), PID2019-110165GB-I00 and PID2022-139427NB-I00 financed by the Spanish MCIN/AEI/10.13039/501100011033/FEDER, UE,  from the Generalitat de Catalunya under contract 2021 SGR 171, as well as by the EU STRONG-2020 project, under the program  H2020-INFRAIA-2018-1 grant agreement no. 824093,  the CRC-TR 211 'Strong-interaction matter under extreme conditions'- project Nr. 315477589 - TRR 211. 

% BibTeX or Biber users please use (the style is already called in the class, ensure that the "woc.bst" style is in your local directory)
 \bibliography{MESON2023_LauraTolos_paper.bib}

\end{document}